# Exact diagonalisation studies of strongly correlated 2D lattice fermions

Didier Poilblanc

*Laboratoire de Physique Quantique, Université Paul Sabatier, 31062 Toulouse, France*

*The T=0 phase diagram of the planar t–J model with small doping is investigated by exact diagonalisations of square clusters with up to 32 sites. At half-filling, the single hole quasi-particule weight vanishes in the strong correlation limit $J \to 0$ while the quasi-particle mass diverges (like $1/J$). However, the spectral function shows weight distributed over a large energy range ($\sim 7\,t$) and a pronounced structure in momentum space which supports the composite particle picture of the hole. For $J/t > J/t|_{B,2}$ the effective attractive force between holes leads to bound-pairs formation. The hole-hole and hole pair-hole pair binding energies calculated on the $\sqrt{26} \times \sqrt{26}$ cluster indicates that in the range $J/t \in [0.16, 0.45]$ a phase of separate pairs of d-wave internal symmetry is stable. On the other hand, the d-wave pair spectral function exhibits a quasi-particle peak down to very small $J/t$ ratios suggesting that $J/t|_{B,2}$ could in fact be as small as 0.05. Above $J/t|_{B,4} \sim 0.45$ the 2-holes pairs bind into 4-holes pairs. An abrupt change of the orbital symmetry of the 4-hole droplet is also observed above $J/t \sim 2.7$. In the parameter range, $J/t|_{B,2} < J/t < J/t|_{B,4}$, Bose condensation of the individual pairs is expected to lead to d-wave superconductivity.*

## 1. INTRODUCTION

### 1.1. Strong Coupling Hamiltonian

Soon after the discovery of superconductivity in the cuprate materials P.W. Anderson[1] and others suggested that this phenomena was intimately linked to magnetism. The t-J model,

$$H = \sum_{\mathbf{x},\mathbf{y},\sigma} t_{\mathbf{xy}} \, c^\dagger_{\mathbf{x},\sigma} c_{\mathbf{y},\sigma} + \sum_{\mathbf{x},\mathbf{y}} J_{\mathbf{xy}} \, \mathbf{S}_{\mathbf{x}} \cdot \mathbf{S}_{\mathbf{y}}, \qquad (1)$$

defined on a two-dimensional (2D) square lattice, emerged as one of the simplest model to describe the low energy physics. The *hole* hopping matrix elements $t_{\mathbf{xy}}$ and the magnetic exchange couplings $J_{\mathbf{xy}}$ (of magnitude t=1 and J respectively) connect only nearest neighbor sites. Hamiltonian (1) is a strong-coupling version of

D. Poilblanc

the well-known Hubbard model. It can also be derived from the more complicated $CuO_2$ Hamiltonian in some relevant parameter regime.[2] In the following we shall present the state-of-the-art of the numerical work on the 2D t–J model.

Table 1
Sizes of the full and reduced Hilbert spaces (corresponding to the ground-state) for various N-sites t–J clusters with $N_h$ holes.

| $N_h / \sqrt{N} \times \sqrt{N}$ | Full Hilbert space | Symmetry group | Reduced Hilbert space |
|---|---|---|---|
| $1 / 4 \times 4$ | 102 960 | $T_{16}$ | 6 435 |
| $1 / \sqrt{26} \times \sqrt{26}$ | 135 207 800 | $T_{26}$ | 5 200 300 |
| $1 / \sqrt{32} \times \sqrt{32}$ | 9 617 286 240 | $T_{32} \otimes C_{4v}$ | 37 596 701 |
| $4 / \sqrt{26} \times \sqrt{26}$ | 10 546 208 400 | $T_{26} \otimes C_4 \otimes I_2$ | 50 717 244 |

### 1.2. Exact Diagonalisation Technique

The Exact Diagonalisation (ED) technique is based on the fact that lattice Hamiltonians like (1) defined on N-sites clusters (with $N_\sigma$ spin-$\sigma$ electrons) are finite *sparse* matrices. Obvious conservation laws allow one to work at fixed magnetization $N_\uparrow - N_\downarrow$ (chosen to be minimum) and number of holes $N_h = N - N_\uparrow - N_\downarrow$. Although the full Hilbert space of size $\mathcal{N} = N!/(N_\uparrow! N_\downarrow! N_h!)$ grows exponentially fast with the number of sites (see Table 1), the number of matrix elements increases only as $N \times \mathcal{N}$ instead of $\mathcal{N}^2$. The Lanczos algorithm which requires a small number of recursive applications of the Hamiltonian matrix can then take advantage of its sparseness. To minimize finite size effects the clusters are chosen with the symmetry of the lattice (square $\sqrt{N} \times \sqrt{N}$ toruses with periodic boundary conditions). As shown in Table 1 use of the symmetry groups (translation group $T_N$, point groups $C_4$ or $C_{4v}$, and spin inversion $I_2$) allow one to conveniently split the matrix into several blocks to be handled separately. ED have been extensively applied to the study of both static and dynamical correlations.[3]

Fig. 1. Single hole and d-wave pair QP weights vs J (from Refs. 4, 10 and 12).

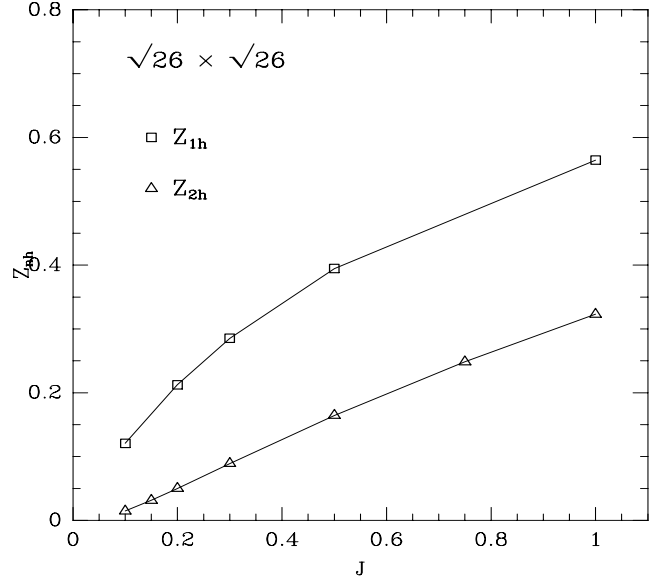

## 2. SINGLE HOLE IN THE ANTIFERROMAGNET

At half-filling the system develops long range antiferromagnetic (AF) order with low energy spin-wave excitations. Upon doping it becomes a metal with a Drude weight proportional to the hole doping.[3] Many studies were devoted to the single hole problem[3] which captures most of the physics of the low doping limit. In this case, it becomes possible to perform finite size scaling.[4,5] These studies show consistently (i) a quasi-particle (QP) band at low energy (see Fig. 1) and (ii) a substantial fraction of the spectral weight of the single hole propagator at higher energy. The minimum of the QP dispersion occurs at the point $\Sigma$ ($\mathbf{q} = (\pi/2, \pi/2)$) and is almost degenerate with the X point ($\mathbf{q} = (\pi, 0)$)[4-7] in surprisingly good agreement with the simple Hartree-Fock treatment. The bandwidth is reduced to $\sim 2.2\,J$.[4,5] In other words the QP mass is enhanced by the correlations.

Fig. 2. Single hole spectral function at momenta (0,0), $(\pi,\pi)$ and $(\pi/2,\pi/2)$ on 32- and 26-sites clusters (from Ref. 10).

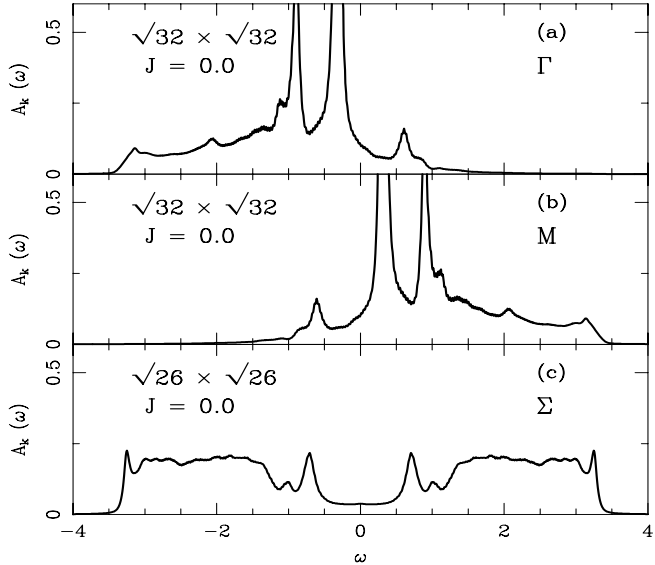

Contrary to one-dimension (1D), a string-like force exists in 2D between the hole charge and the hole spin centers of mass.[8] However, finite size scaling[4] shows that the QP weight desappears as $\sim J^{0.6}$ when $J \to 0$ (see Fig. 1) suggesting that the hole eventually breaks apart in this limit.[9] On can then interpret the QP as a bound state between two different objects, a light one (with a mass $\sim 1/t$) and a heavy one (with a mass $\sim 1/J$) in analogy with the 1D case.[9] The confining string-like force vanishes in the $J \to 0$ limit where the two components become eventually free. This picture implies then naturally that *simultaneously* (i) the QP mass diverges when $J \to 0$[3,4] and (ii) the spectral function (see Fig. 2) exhibits weight over an energy scale of several t as well as some well-defined structure in **k**-space.[5,10] Moreover, this is also consistent with the fact that the optical mass (extracted from the Drude peak) $\propto 1/t$ and the QP mass $\propto 1/J$ are essentially different.[11,5]

Fig. 3. $\Delta_4$ for various symmetries of the 4-hole GS. The short-dashed and long-dashed lines correspond to the exact large-J extrapolation and to the estimation based on the exact Heisenberg magnetic bond energy respectively (from Ref. 18).

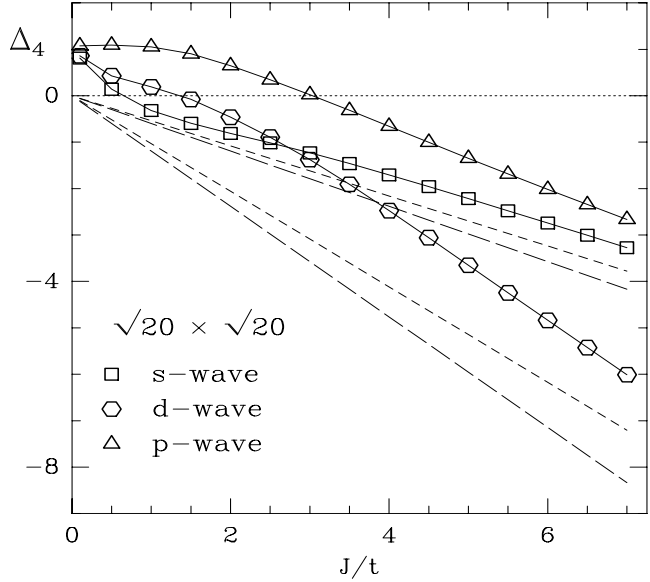

## 3. BOUND PAIRS FORMATION

For increasing J, the QP pole becomes stronger. A simple broken-bond counting argument shows that, at large J/t, two holes injected in the AF can minimize the local magnetic energy by forming a bound state. Finite size scaling have shown[12] that this picture is actually correct even down to $J = J_{B,2} \simeq 0.2$ (see also Fig. 4). The hole-hole pair has an $d_{x^2-y^2}$ orbital symmetry although it exhibits dominant hole-hole correlations at an intermediate distance of $\sqrt{2}$ ie when the holes stay across the diagonal of a plaquette on the *same sublattice*.[13] The two-hole bound-state also exhibits flux quantization in a ring in units of $hc/2e$.[11]

When pairs are formed in this parameter regime they are likely to Bose condensate into a superconducting state unless some sort of clustering occurs.[14] However, early ED[15] as well as high temperature expansions[16] seem to consistently find that phase separation (PS) occurs only for $J \geq J_{PS} > J_{B,2}$. The effective binding energy between pairs can be quantitatively defined in the low doping limit by

Exact Diagonalisations of 2D Correlated Fermions

Fig. 4. Binding energy $\Delta_4$ vs J for clusters of 16, 20 and 26 sites (from Ref. 18). The 26-sites cluster hole-hole binding energy $\Delta_2$ is also indicated by open stars.

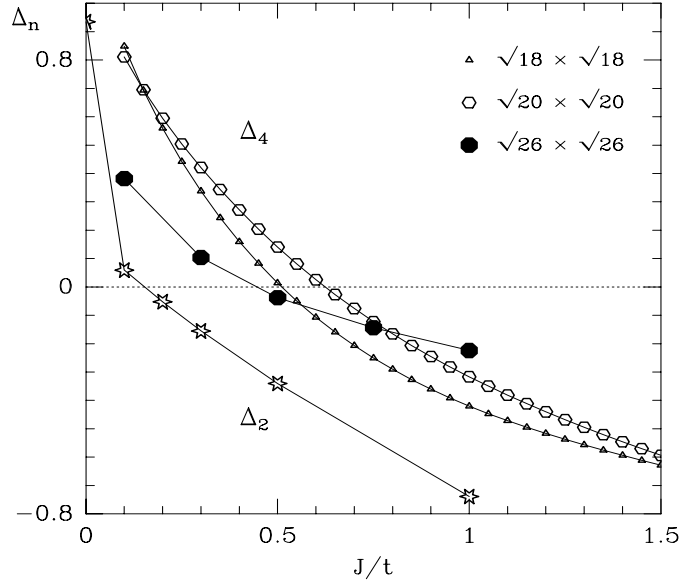

$\Delta_4 = E_{4h} + E_{0h} - 2E_{2h}$ where $E_{nh}$ is the ground-state (GS) total energy for a system with $N_h = n$. GS energies of 4-holes can be calculated for the three possible orbital symmetries of the wave function (with zero total momentum) and the corresponding data for $\Delta_4$ are shown in Fig. 3. The preformed pairs bind (ie $\Delta_4 < 0$) when J exceeds some critical value which depends on the orbital symmetry. The physical character of each wave function can be identified from the magnetic energy lost in the large J limit (see Fig. 3). A *first order* transition is seen around $J \simeq 2.7$ between a state with lines of holes to a state with holes forming a real droplet. The first order character of the transition in fact explains the rapid change in the hole-hole correlations observed recently by approximate methods.[17] In order to estimate the critical J $\Delta_4$ has been evaluated on various lattices of size up to N=26 and the data are displayed in Fig. 4 and Table 2. Note that the critical value $J_{B,4}$ at which $\Delta_4$ changes sign depends weakly on the system size while the slope $|\partial \Delta_4 / \partial J|$ at this point decreases for increasing size. Fig. 4 then strongly suggests that if $J/t \in [J/t|_{B,2}, J/t|_{B,4}]$ with $J/t|_{B,2} \sim 0.16$ and $J/t|_{B,4} \sim 0.45$ individual pairs exist without forming larger clusters of holes. Note however that $J/t|_{B,2}$ for pair formation could in fact be smaller as suggested by the observation that the dynamical response of the d-wave pair operator exhibits on small clusters a sharp

Table 2
Binding energies $\Delta_2$ and $\Delta_4$ calculated on a 26-site cluster.

| J/t | 0.1 | 0.3 | 0.5 | 1 |
|---|---|---|---|---|
| $\Delta_2$ | 0.05953 | -0.15496 | -0.34020 | -0.73861 |
| $\Delta_4$ | 0.38198 | 0.10378 | -0.03776 | -0.22412 |

## D. Poilblanc

$\delta$-peak of weight $Z_{2h} \propto J/t$ down to very small values of $J/t$[12] as shown in Fig. 1. The formation of droplets should not be confused with real phase separation. More insight could be obtained from the calculation of the inverse compressibility $\kappa^{-1}$ defined by $\kappa^{-1} = \partial^2(E_{GS}/N)/\partial\langle n_h\rangle^2$. We expect $\kappa^{-1}$ to drop discontinuously to zero in the phase separated regime. A non-uniform phase with domains walls could also compete with phase separation. A careful finite size scaling is necessary to investigate such possibilities and the issue of phase separation remains certainly an open question.

## ACKNOWLEDGMENTS

This research is supported by C.N.R.S., France. The computer simulations were done on the CRAY-C98 at I.D.R.I.S, Orsay, France. I thank W. Hanke, F. Mila, R.G. Laughlin, D.J. Scalapino and T.L. Ziman for stimulating discussions.